# Guided Support for Collaborative Modeling, Enactment and Simulation of Software Development Processes

Ines Grützner, Jürgen Münch
*Fraunhofer Institute for Experimental Software Engineering*
*Sauerwiesen 6*
*67661 Kaiserslautern*

*Phone:* *+49 (0) 6301 707 0*
*Fax:* *+49 (0) 6301 707 100*
*Email:*[*Ines.Gruetzner|Juergen.Muench@iese.fraunhofer.de*](mailto:Ines.Gruetzner|Juergen.Muench@iese.fraunhofer.de)

Alejandro Fernandez, Badie Garzaldeen
*Fraunhofer Institute for Integrated Publication and Information Systems*

*Phone:* *+49 (0) 6151 869 4837*
*Fax:* *+49 (0) 151 869 963*
*Email:* *Casco|Badie.Garzaldeen@ipsi.fhg.de*

## Abstract

*Recently, the awareness of the importance of distributed software development has been growing in the software engineering community. Economic constraints, more and more outsourcing of development activities, and the increasing spatial distribution of companies come along with challenges of how to organize distributed development.*

*In this article, we reason that a common process understanding is mandatory for successful distributed development. Integrated process planning, guidance and enactment are seen as enabling technologies to reach a unique process view.*

*We sketch a synthesis of the software process modeling environment SPEARMINT and the XCHIPS system for web-based process support. Hereby, planners and developers are provided with collaborative planning and enactment support and advanced process guidance via electronic process guides (EPGs). We describe the usage of this integrated environment by using a case study for the development of a learning system.*

## 1. Introduction

Nowadays, more and more software development projects are performed by teams that are distributed across several locations. Outsourcing of development activities and time-to-market pressure, especially enforce the spatially as well as temporally distributed planning and enactment of software processes. Distributed teams need detailed guidance on the development process to be followed in the project and comprehensive support in planning and enacting this process in order to finish their project successfully. One main project success factor provided by detailed process guidance and comprehensive enactment support is the communication of the development process among the team members. If team members know the process to be followed, they can identify an adequate process for the project during project planning, can coordinate their work and know, who does which activity, when, and with whom. For example, tasks will not be done twice by different team members, are less error prone, or will not be forgotten at all. For these reasons, it is expected that integrated planning, enactment and guidance support will lead to a tremendous reduction in project effort.

To provide detailed process guidance and comprehensive process planning and enactment support, an environment is required that, for example,

- supports process elicitation, modeling, and guidance in order to model and follow the development process of a project (R1),
- supports collaboration (e.g., negotiation) of team members during the tailoring of the given development process to their project in order to apply the adequate process (R2),
- supports further tailoring of the process during execution in order to cope with changing contexts and requirements or react to emerging problems (R3),
- combines executable process models with detailed process guidance and resources, for example, templates and examples of work products in order to provide a single work environment for all team members with process guidance, process information, and a workspace to work on documents collaboratively (R4),
- supports improvement of the development process while learning from the practice (R5), and
- supports "testing" of process behavior before implementation by means of simulation (R6).

A comprehensive list of requirements for flexible process modeling and enactment environments can be found in [1].

This article describes a web-based process modeling, enactment, and simulation environment that fulfills the requirements mentioned above. It consists of the SPEARMINT process modeling environment and the XCHIPS system for web-based process support that are integrated via an XML interface for process model exchange and online-guidance interfaces. In Section 2, the process modeling environment SPEARMINT is introduced. Spearmint allows for the generation of web-based guidance support. Section 3 describes the XCHIPS environment that addes further collaboration and enactment functionality. In Section 4, the integration of both parts is demonstrated using an example from the project "e-Qualification Framework (e-QF)". The aim of the e-QF project was to develop an innovative learning environment (including the web-based process modeling, enactment, and simulation environment), which is amended with methodologies supporting authors in the production of courseware for the environment. Finally, initial experience with the integrated environment and directions for future work are outlined.

## 2. Process Guidance with the SPEARMINT Process Modeling Environment

SPEARMINT/EPG [2,3,4] aims at supporting the modeling and online-documentation of software development processes from a software engineering perspective. Properties of such processes are, for example: many people are involved in a project and perform many different types of tasks; the processes are complex and abstraction techniques are needed to model them; not all process steps are known in advance when planning the project and changes in the model in the online-documentation are often necessary. One of the main ideas of SPEARMINT/EPG is to split descriptions of software development processes into views. Each view is generally defined as a projection of a process model that focuses on selected features of the process.

A graphical notation is used to describe development processes. It distinguishes between activities, artifacts, roles, tools, and their attributes, which correspond to measurable qualities of the objects.

Traditional process descriptions are typically documented in handbooks or, in the scientific community, using formal notations (such as process modeling languages). The latter usually requires a transformation in a graphical representation so that it can be used it practice. The use of handbooks in the software development process has been recognized widely as beneficial in order to perform systematic and traceable local projects. Nevertheless, when using process handbooks software developers face problems which are caused by the informal style of representing those descriptions and the difficulty of maintaining them consistently. The informal description of software engineering processes in handbooks results in problems in their usage:

- Process descriptions in handbooks are lengthy, perhaps hundreds of pages, and often not very well structured thus making information retrieval difficult.
- Handbooks often lack role-specific views which makes it difficult for project members to play particular roles to find relevant information with respect to their specific problems. Views can also be an important means for providing relevant process knowledge to several development sites that are distributed.
- It is difficult to modify informal process descriptions. This aggravates the adoption of the processes to the organizational contexts in which they are used. Tailoring to particular project characteristics and goals is needed.

- The dynamic behavior of a process is difficult to understand. Descriptions in process handbooks are unsuited to serve as input for simulators or execution machines.
- The consistency, unambiguity, and completeness of software process descriptions cannot be ensured on an informal basis. Costly reviews are needed to develop a high-quality process descriptions.

Computer support for managing large amounts of process knowledge is desirable. Electronic process guides (EPGs) could be used to navigate through the information space; filters would provide only meaningful data to particular roles; a formal internal representation of the models languages would enable checking of the models. These aids would allow a development project to be performed more efficiently and effectively by providing well-defined environments which support various roles in a project.

Spearmint/EPG is such an environment for managing large process descriptions. In particular, it allows for the generation of so-called Electronic Process Guides (EPGs) [4], which are generated Internet/intranet hypertext documentations of the process information. The purpose of EPGs is to guide software developers in doing their tasks by providing the relevant information they need (e.g., process descriptions, links to documents such as checklists). One of the main benefits of SPEARMINT/EPG is that it is based on a maintainable XML-based process representation. Changes of the process models can easily be propagated to different views and also to the online-documentation (because it is generated). Another important benefit of these EPGs is that they support distributed process planning by providing the appropriate representations for reviews. The combination of EPGs with process engines can be performed by generating appropriate links from objects of the enactment environment to EPG fragments. Besides the advantage, that SPEARMINT process description allow for easy maintenance and generation of EPGs, two other benefits can be seen:

1) SPEARMINT process models are appropriate means for storing software development knowledge. In general, reusing experience (e.g., process models) is a key to systematic and disciplined software engineering. Although there are some successful approaches to software product reuse (e.g., class libraries) improvement should comprise the reuse of all kinds of software-related experience, especially process-related experience. SPEARMINT process models can be used as initial assets of an experience repository. SPEARMINT process models are means for capturing the relevant aspects and can be stored using various structures of an experience repository (e.g., type hierarchies, clusters of domain specific assets).

2) SPEARMINT allows several kinds of (automated) analyses, which can be performed before the project starts, during process enactment and in a post-mortem fashion after project termination. Process models can, for instance, be analyzed statically (e. g., with consistency checking) and compared with each other. The latter is for example important during the modeling of the interfaces of distributed processes.

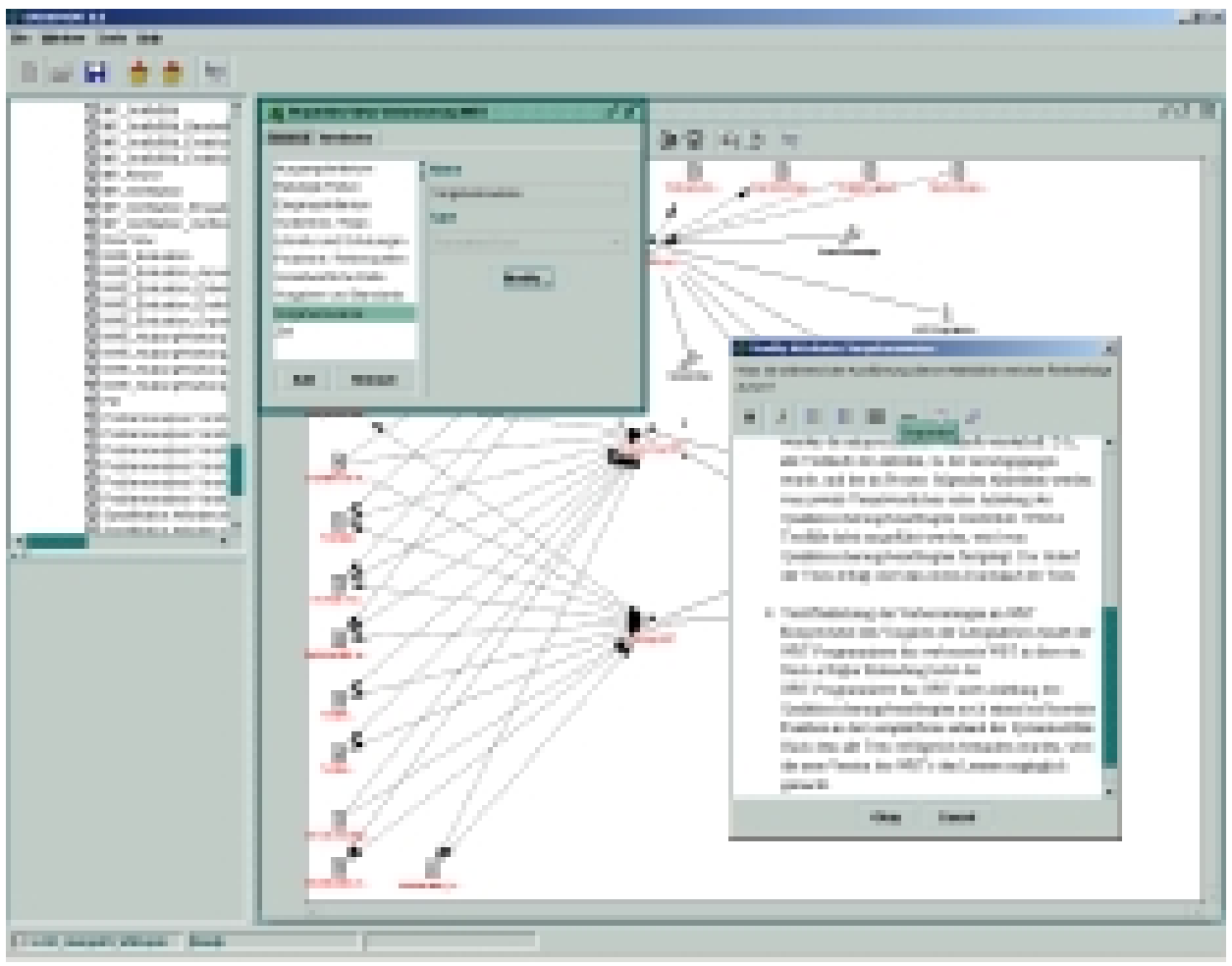

**Figure 1. The SPEARMINT user interface**

These described aspects (easy maintenance, EPG generation, capturing process knowledge for reuse, and analysis capabilities), lead to the decision to use SPEARMINT as the modeling tool in the process environment.

## 3. Process Enactment with XCHIPS

The XCHIPS system (XCHIPS stands for Cooperative Hypermedia Integrated with Process Support) supports collaborative modeling, tailoring and enactment of work processes. Process models are constructed on the basis of a graphical, hypermedia structure [5].

A unique feature of XCHIPS is that all changes to a process model during modeling or enactment can be made in collaboration. That is, many users can at any time access a process model concurrently. Consistency, and synchronous replication of changes are ensured. This, together with annotation and other cooperation facilities, makes XCHIPS a powerful tool to support synchronous collaboration between distributed users who need to negotiate and change their work processes.

A process model in XCHIPS is based on a predefined modeling language. This language states which types of elements and links are available and which restrictions for composition and connection apply. A meta-modeling tool [6] that accompanies XCHIPS allows tailoring the language to a specific use. This is valuable, for example, when integrating XCHIPS with other process modeling and management tools and methods.

XCHIPS process models can be enacted. Similarly to workflow systems, the enactment engine ensures process restrictions (e.g. dependencies), and resolves document flows. However, XCHIPS process models can be tailored during execution, which copes with the changing contexts of software development processes. It is also possible to enact incomplete process models. The missing parts can be completed as needed. This is a powerful feature when supporting software projects where the plans cannot be totally defined before the project starts.

XCHIPS is implemented in pure Java and deployed using web technologies (i.e., Java plugin and Java Webstart). As a tool to support collaboration online, XCHIPS can be configured with extra collaboration functionality such as videoconference, shared whiteboard, integrated chat, and a shared notepad. These tools can also be included as elements in process models to indicate points in the process where collaboration needs to take place.

## 4. Integration of SPEARMINT and XCHIPS into a Web-based Process Modeling, Enactment, and Simulation Environment

The integration of the SPEARMINT process modeling environment and the XCHIPS system by an XML interface to exchange process models results in a web-based process modeling, enactment, and simulation environment. This environment, which is part of the learning environment developed in the e-QF project, aims at meeting the requirements of distributed collaborative work mentioned at the beginning of this abstract. It guides and supports courseware authors following the development process of the IntView courseware engineering methodology [7] in producing courseware for the e-QF learning environment and the development of appropriate software/web support.

In the following, we will explain in more detail how the requirements are realized using a scenario in the context of the e-QF project. During this project, the courseware "Process modeling, planning, and enacting with SPEARMINT and XCHIPS" for use in the e-QF learning environment is developed by a distributed team. This team is supported by the web-based process modeling, enactment, and simulation environment.

The prerequisite for running the project is the explicit modeling of the development process of the IntView courseware engineering methodology with SPEARMINT (R1). The SPEARMINT environment was chosen for modeling the development process because of its sophisticated, easy to use process modeling features introduced in chapter 2 and of its abilities to generate an EPG. These features allows for developing a comprehensive process model with a corresponding EPG, which could not be developed with the still rudimentary process modeling features of the XCHIPS system.

The SPEARMINT process model of the IntView courseware engineering methodology contains detailed descriptions of each activity and each artifact developed during the production of the courseware by the means of SPEARMINT attributes as well as examples and templates of these artifacts. It also provides an introduction to the methods or tools supporting the enactment of the activities, and to the roles required to perform the modeled process. Furthermore, a small experience base of useful hints, of guidelines and standards, as well as of problems in performing the activities is integrated into the process model using SPEARMINT attributes.

The SPEARMINT process model also provides the control flows, which are required by the XCHIPS system in order to establish the activity flow in the XCHIPS templates.

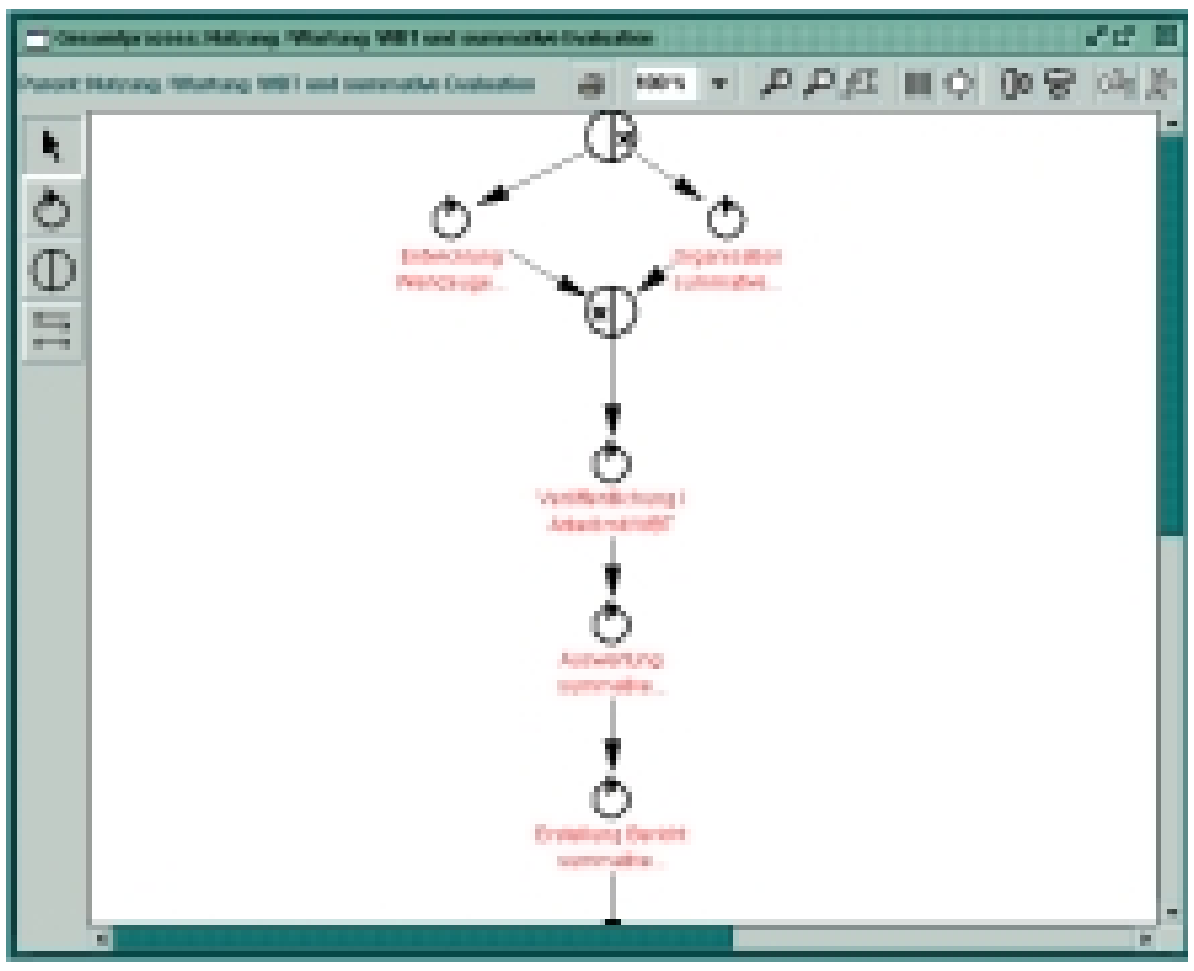

**Figure 2. A sample control flow of the IntView process model in SPEARMINT**

After modeling, the EPG [8] is generated from the SPEARMINT process model (see Figure 3). The process model is also exported as XML model ready to be imported by the XCHIPS system. Both the XML model and the EPG are static models. They are static in the sense that they cannot be instantiated for particular occurrences of the project and that they do not provide support for execution. Therefore, the XML model of the IntView development process has to be imported into the XCHIPS system. By importing the XML model into XCHIPS, one obtains a living process template that can be tailored and instantiated for specific projects.

The XCHPS system provides a service for the import of the XML model and for the integration of the associated EPG with the help of links from objects imported to the appropriate EPG pages. The graphical user interface and the hypermedia support of XCHIPS simplify navigation in the imported process model. The links from the graphical presentation of objects in the process model to their specification in the EPG give a context-sensitive guide.

A template browser provides access to all templates that are currently available in the XCHIPS system. The browser also provides the user with a list of the all projects that are currently being planned, and with a list of all projects that are currently running (i.e., instances). But a user has write access only to the projects in which he/she involved in. The access rights are granted by the initiator of the project or later by other some team member. Using the browser, distributed users can collaboratively browse, annotate, tailor and clone the templates in order to initialize and tailor the processes behind the templates to their projects. For example, the project to produce the WBT "Process modeling, planning, and enacting with SPEARMINT and XCHIPS" is planned collaboratively by the project managers from both locations of the distributed team. During planning, these project managers meet online to establish a project plan on the basis of the template for the IntView development process. The managers can also have a look at existing instances of the template (e.g., finished projects) to learn from past experiences.

The goal of the planning phase is to tailor the project. This includes completing the template with all the necessary details that are needed before starting execution, namely the persons that will fill the different required roles, and any notes or documents that may be needed. To start the planning, the project managers select the IntView template and clone it. The clone is then placed in the list of projects being planned. Then, the project managers give a name to the template and open it. They tailor it to get their project-specific process and the right assignment of team members to each activity of the project-specific development process (R2). They do it in synchronous online collaboration, using a chat tool for their communication. The log of the chat can be kept in the context of the discussion by embedding it as an element into the process model.

While planning the project to develop the WBT, the enactment of the tailored process can be simulated in XCHIPS in a type of synchronous role-playing (R6). Synchronous role-playing means running the project in a time-lapse mode but not simulating it by means of using dynamic or discrete simulation models. That is, each participant takes a role in the project (possibly more than one if one of the assigned persons is not on-line) and the project is started. Participants check for each of the tasks they are assigned to that the preconditions are set correctly, that resources are available, and that the needed guides are available. If this is the case, they mark the task as being finished. This causes the execution engine to move process forward. This form of role-playing helps to ensure that required documents flow correctly, and that dependencies and therefore task activation behave as expected.

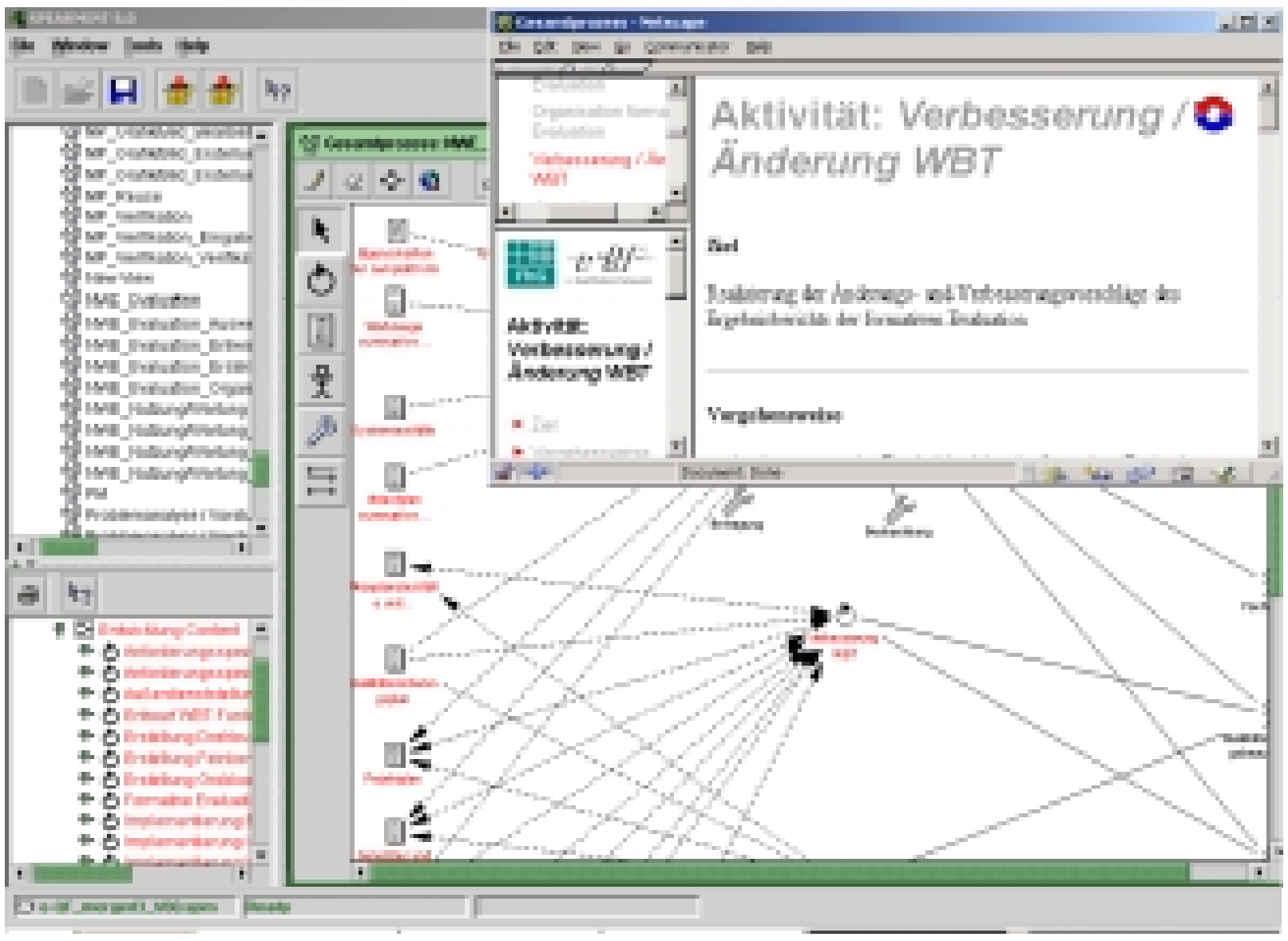

**Figure 3. Part of the IntView process model in SPEARMINT with the corresponding EPG**

Role-playing also has an important role in training participants before the actual project starts. Figure 4 shows a project being simulated. A composed task has been opened in a new window. The following task states are distinguished: 'finished', 'enabled', 'inactive', and 'active'.

As soon as planning of the early phases of the courseware development project is complete, the project can start. To do this, one of the project managers selects the project from the list of projects being planned and starts it. This action moves the project to the list of running projects. If the start task is modeled, it is automatically activated, started and finished. This starts the chain of activations of all initial tasks (that had start as a prerequisite). If there is no start task indicated, some user will have to manually activate all initial tasks. If there is a notification component assigned to the active initial tasks, an email notification about the activation of each task is sent to the team members assigned to this active task.

When a team member receives an email notification about the start of a task he/she is assigned to, he/she logs in to XCHIPS and selects the WBT development project (indicated in the email) in the list of running projects. With the help of the search tool (see Figure 5), he/she locates the task he/she is assigned to. In order to get a more detailed specification of the work to be done in this task, he/she can open the corresponding entry in the EPG directly from the XCHIPS task (R4). This reference entry contains all details about the task (including the product flow in order to establish communication between the different tasks), about the artifacts to be produced, and about available methods/tools for performing the task. in the process description and get a more global view of the activities to be performed.

Furthermore, the team member gets access to templates and examples of the artifacts to be produced in the activity.

In order to enable learning from practice and to support the capture of the experience made by the distributed team during the course of the project (R5), the EPG can be extended by an annotation feature [9]. When writing an annotation, each team member can assign his/her experience with a specific process model element directly to this element. The project managers can use these annotations to smooth the development process (that is, re-plan the current project or optimize the plan of a new project) or to change, adapt, or improve the IntView development process itself. For the same purposes, it is planned to implement a mechanism that allows to export tailored processes as SPEARMINT compatible XML models. These XML models are to be re-imported into the SPEARMINT environment in order to be the basis of improved IntView process models.

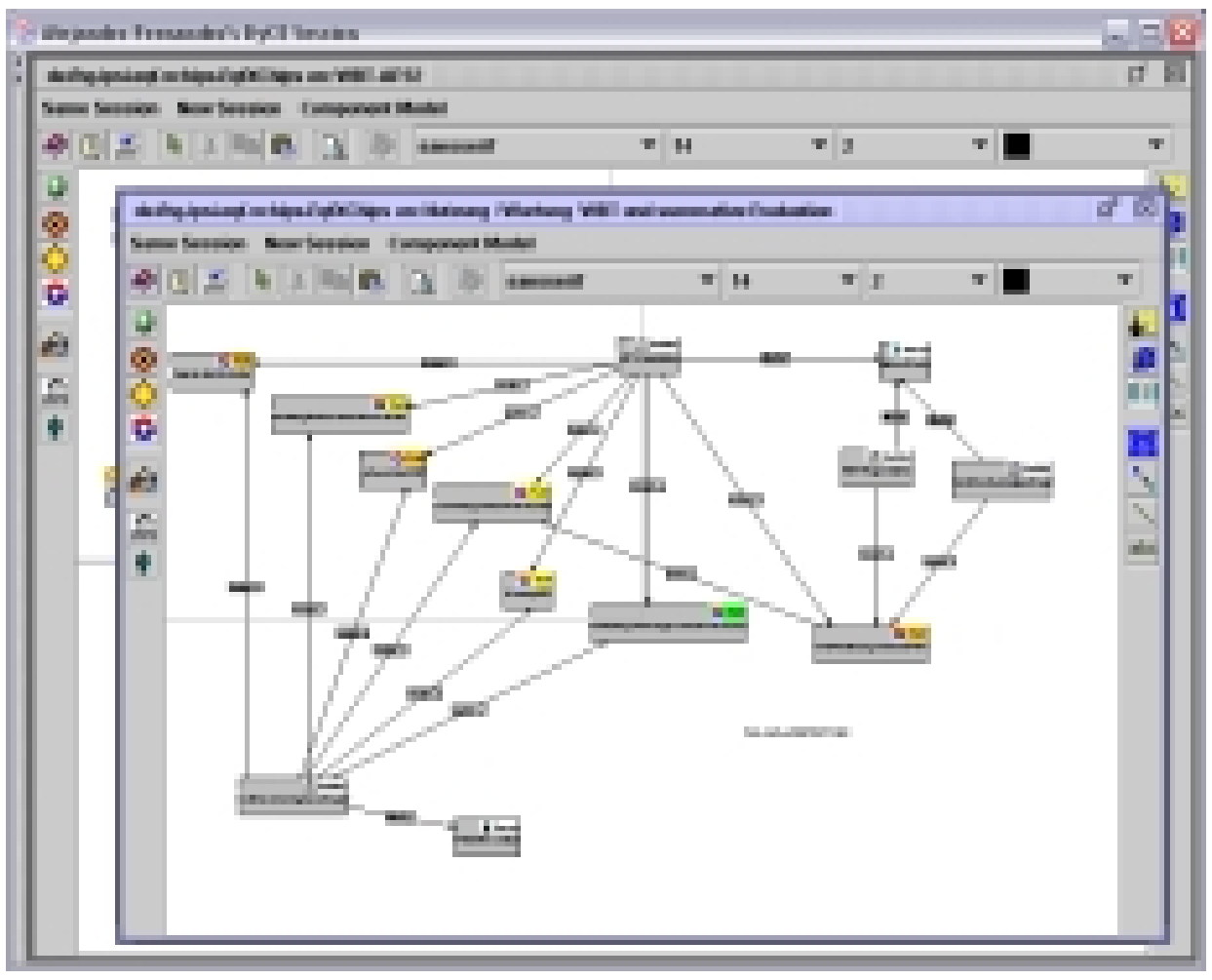

**Figure 4. A process model of a subtask in a running project.**

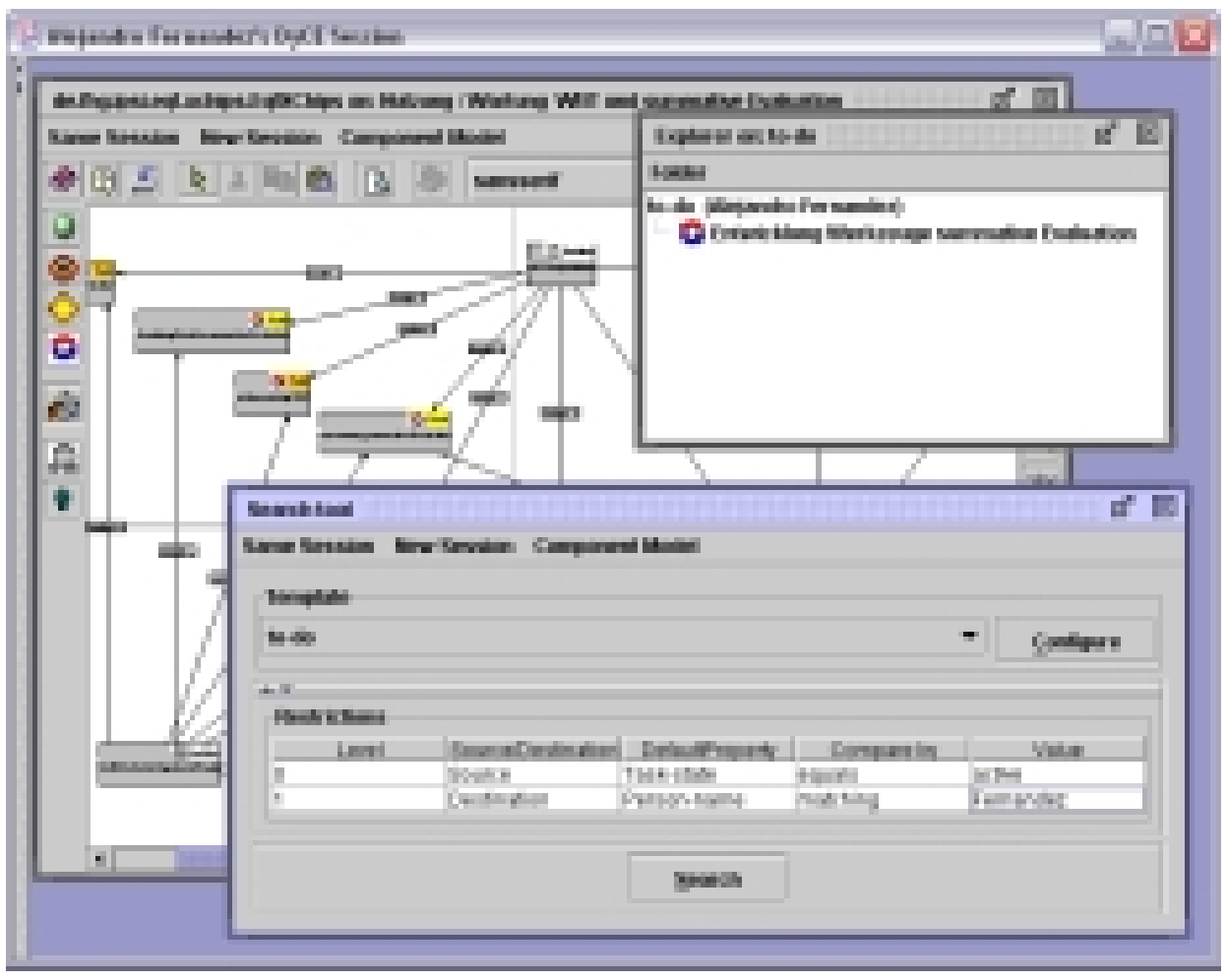

**Figure 5. Searching for tasks in a XCHIPS running project**

## 5. Conclusions

First scenario-based evaluations of the operative prototype of the integrated environment have been performed with distributed participants from the application domain in the course of the e-QF project. Experience shows that the combination of the XCHIPS process model and the EPG is of great value for project participants who did not yet have any experience with the IntView methodology. They indicated that they found the tips and detailed descriptions in the guide a valuable source for learning about their role in the project. The participants also welcomed the possibility of simulating the execution of a process in a kind of role play. However, during the first usage experiences, synchronous interaction did not occur often because the participants were not used to the possibility of working synchronously on shared documents. As the participants become more and more aware of this opportunity and get accustomed to it, synchronous modeling and negotiation sessions will become more common. The prototype can be demonstrated at the ProSim 2003 conference.

The work led to several new research questions, i.e., how to support distributed development from the management point of view? Which forms of interconnected global organizations (e. g., virtual cooperations, network organizations, global learning organizations) require which kind of process support? Traditional management hierarchies are to be replaced by organizational structures that allow to distribute power according to who has the relevant resources, information and capabilities to contribute to the task at hand. New delegation and decision processes are needed as well as the integration of different globally distributed development processes. An integrated conceptual framework for handling all “global” aspects of software processes should be investigated in the future.

## Acknowledgements

The development of the IntView methodology and the XCHIPS environment was partly funded by the “e-Qualification Framework (e-QF)” project under grant 01AK908A of the German Federal Ministry of Education and Research (BMBF). The integration of SPEARMINT with simulation is supported in part by the German Federal Ministry of Education and Research (SEV project, grant 01AK943A) and the Stiftung Rheinland-Pfalz für Innovation (ProSim Project, no.: 559).